\documentclass[preprint,showpacs,preprintnumbers,displaymath,amsmath,amssymb]{revtex4}

\usepackage{graphicx}
\usepackage{verbatim}
\usepackage{color}
\usepackage{subfigure}  
\usepackage{dcolumn}
\usepackage{bm}
\usepackage[dvips]{epsfig}
\usepackage{bbm}
\usefont{T1}{tnr}{m}{sl}
\newcommand{\mf}{\mbox{\boldmath$\mu$}}
\newcommand{\mbf}{\mbox{\boldmath$\mu_{1}$}}
\newcommand{\mcf}{\mbox{\boldmath$\mu_{2}$}}
\newcommand{\kbf}{\mbox{\boldmath$\kappa$}}
\newcommand{\rbf}{\mbox{\boldmath$\rho$}}
\newcommand{\Rbf}{\mbox{\boldmath$R$}}

\begin{document}

\title[c]{Quantum well states in Fe/Nb(001) multilayers: First principles study }

\author{Nitya Nath Shukla}
\altaddress{Current Address: Electronic Research Administration, National Institutes of Health (NIH), 6705 Rockledge Dr., Bethesda, Maryland 20892, USA}
\author{A. Sen}
\altaddress{Current Address: Condensed Matter Theory Division, Research Center for Applied Sciences, Academia Sinica, Taipei 11529, Taiwan}
\author{R. Prasad}
\email{rprasad@iitk.ac.in}
\affiliation{Department of Physics, Indian Institute of Technology, Kanpur 208016, India}

\begin{abstract}
We present a $first$-$principles$ study to understand the phenomena of interlayer exchange coupling in Fe/Nb multilayers using the linearized-muffin-tin-orbitals method within the generalized gradient approximation. We find that the exchange coupling oscillates with both short and long periodicities, which have been examined in terms of the Ruderman-Kittel-Kasuya-Yosida (RKKY) model as well as the quantum well (QW) model. We have investigated the behavior of the exchange coupling by artificially varying moments of Fe atoms in ferromagnetic layers. For small moment of Fe, the coupling shows bilinearity in the magnetic moments implying its RKKY character. However, at higher moments close to the bulk Fe, the saturation of long-period oscillations is in accordance with the QW model. Quantum-well dispersions around the Fermi level demonstrate that the majority-spin bands contribute largely to the formation of quantum-well states, which we analyze quantitatively by making use of the phase accumulation model. Our analysis indicates that the quantum well model gives a better 
description of the oscillatory behavior of the exchange coupling in Fe/Nb multilayers.
  
\end{abstract}


\preprint{MANUSCRIPT}
\maketitle

\section{Introduction}
Damped long-range oscillation of the interlayer exchange coupling (IEC) as a function of the spacer thickness is a well known phenomenon in magnetic multilayers\cite{pg,ssp,mn}. Several approaches have been proposed over the years in order to explain the oscillatory behavior of the IEC. Among these, the two prominent models are (i) the Ruderman-Kittel-Kasuya-Yosida (RKKY) model\cite{bruno} and (ii) the quantum well (QW) model\cite{dm}. The RKKY interaction stems from the spin-polarization of the intervening conduction electrons in the spacer layer. Bruno and Chappert\cite{chappert} have shown that the exchange coupling within the RKKY theory is related to the topological properties of the Fermi surface of the spacer material. Since the magnetic atoms in multilayers are more immersed in the spin sea of the conduction electrons of the spacer layer near the interface, the RKKY interaction has the strongest effect at the interfaces\cite{baj}.   
However, the RKKY approach is not much effective\cite{bruno,chappert} in describing the correct $amplitude$ and $phase$ of the coupling oscillations essentially because the interaction between the ferromagnetic layers and the conduction electrons is not well captured in the RKKY approach. In the Quantum Well model, the coupling arises due to the spin-dependent confinement\cite{mvs} of electrons inside the spacer medium as the size of the multilayer system is reduced to the nanometer range. In the quantum well picture, each layer thickness of magnetic as well as nonmagnetic (NM) kind in the entire multilayer stack contributes significantly to the coupling strength implying that the IEC is not a sheer interfacial effect\cite{qiu}. The QW states are formed near the Fermi level when spin-polarized bands are shifted away from the Fermi surface because of the strong magnetization on both sides of the spacer medium. Such states, which can be observed directly by photoemission measurements, shift in energy with the spacer thickness and become closely spaced when the spacer layer appears to be sufficiently thick. Though both the models arrive at the same period of coupling oscillations due to their origin in the shape of the spacer Fermi surface, they differ in determining the coupling strength\cite{himp2}. It is because the RKKY model originates from the second order perturbation theory whereas the QW model does not make any assumption about the strength of the interlayer interaction. Further, QW theory predicts additional possibilities like oscillating density of states, quantum well dispersions etc., especially when the spacer layer forms a multisheet Fermi surface (FS). Such is the case in Fe/Nb multilayers, where Nb spacer has three sheets of Fermi surface in the (100) plane\cite{gwc}. 

Although many computational works exist already in the literature\cite{ssp,bruno,wang} on the exchange coupling phenomena in systems such as Fe/Cr, Fe/Au, Co/Cu etc, where the spacer layers are in general NM transition metals, not much theoretical understanding has been gained so far for Fe/Nb multilayer system. The first experimental study on the sputtered Fe/Nb superlattices as carried out by Mattson $et$ $al$\cite{matson} reported a weak coupling with an oscillation period of about 9 {\AA}  at room temperature. However, some $ab$ $initio$ band structure studies\cite{sticht,nitya} on  Fe/Nb multilayers demonstrated an oscillation period of 4.6--6.0 {\AA}. Neutron reflectometry data of Rehm $et$ $al$\cite{rehm} for Fe/Nb multilayers have earlier suggested an oscillatory RKKY kind of coupling for small Nb layers. In another experimental development, Klose $et$ $al$\cite{klose} have shown that hydrogen charging can  modify the magnetic coupling in these heterostructures through the alteration of the electronic structure of the Nb interlayer.   

Fe/Nb multilayers seem to be interesting systems to study the interlayer exchange coupling in the sense that Fe is a strong transition-metal ferromagnet (FM) while the spacer layer can be a superconductor (SC) at low temperatures. Thus apart from the phenomena of oscillatory exchange coupling between FM layers, Fe/Nb superlattices have another interesting phenomenon known as the "proximity effect", which paves the way for new sources of magneto-resistance with potential applications in magneto-electronics\cite{lr}. The presence of the internal magnetic field in Fe layers weakens the phenomena of superconductivity due to the breaking of Cooper pairs. Among various FM/SC hetero-structures, where superconductivity gets  induced in the ferromagnet by bringing it in contact with the superconductor, Fe/Nb multilayers have been studied experimentally\cite{rehm,kk,muhge} to an appreciable extent. One such observation by M\"uhge $et$ $al$\cite{muhge} suggests that the thickness of the SC interface mainly determines the actual shape of $T_{c}$ versus $d_{Fe}$ (thickness of the FM layer) curve in Fe/Nb superlattices. Since Nb is the highest known $T_{c}$ element, its interplay with Fe continues to remain an active field in the study of such heterostructures. 

In the present work, we elucidate the coupling phenomena in Fe/Nb multilayers within the domain of the density functional theory\cite{dreizler}. Previous calculation of Shukla and Prasad\cite{nitya} reported the IEC for the Fe/Nb multilayer system up to 7 monolayers of Nb spacer sandwiched between two Fe layers. However, it could not shed light on the detailed analysis of the coupling phenomena due mainly to the lack of relatively large spacer thickness, which is necessary for such kind of study. This work thus involves Fe$_{3}$Nb$_{m}$ (m=1..16) system in order to understand the coupling behavior in Fe/Nb(001) multilayers. Our calculation shows that QW states are indeed formed in such heterostructures and 
the QW model gives a better description of the oscillatory exchange coupling in Fe/Nb multilayers.  

The organization of the paper is as follows. In Sec. II, we briefly outline the computational procedure adopted in the present study. Sec. III, under several subsections, deals mainly with our results that include simultaneous discussions. We finally sum up our observations in Sec. IV.   

\section{Methodology}
Total energy calculations have been carried out for Fe$_{3}$Nb$_{m}$ (m=1..16) multilayers in the framework of the linearized-muffin-tin-orbitals (LMTO) method\cite{ok1,hls,oka} within the tight-binding representation.
The atomic-sphere approximation (ASA) is used for the potentials determined self consistently in the generalized gradient approximation (GGA)\cite{pw} of 
the density functional theory\cite{dreizler}. Various tetragonal supercells are constructed out of Fe and Nb monolayers (ML) where the bcc Fe layers are stacked along the [001] growth direction (see Ref. 15) in ferromagnetic (FM) as well as antiferromagnetic (AFM) orientations. The interlayer exchange coupling (IEC ), denoted by J(m), corresponds to the energy difference between FM and AFM configurations per unit cell structure so that
\begin{equation}
J(m) = E^{\uparrow\downarrow}_{tot}(m) - E^{\uparrow\uparrow}_{tot}(m),
\end{equation}
where $m$ is the number of spacer layers. 
We calculate the total energies of all occupied states and minimize it between the FM and AFM configurations in the self-consistent fashion for each Nb thickness. The average lattice parameter  with reduced lattice mismatch is taken\cite{nitya} as 3.067 {\AA} for the present heterostructures. Linear tetrahedron method has been used for the Brillouin zone (BZ) integration with a maximum of 840 $k$-points in the irreducible wedge of the surface BZ. We use the same unit cell for the FM and AFM structures to obtain reliable energy differences between ferromagnetic and antiferromagnetic ordering. 

\vspace{5mm}
\section{Results and discussion}
\subsection{Interlayer coupling oscillations}
Initially, we compute the IEC as a function of the Nb spacer thickness using Eq. (1) for FeNb$_{m}$($m$ = 1..7) system. The results for $J(m)$ are in good agreement with our earlier calculation\cite{nitya} using the FP-LAPW method. This gives us confidence in carrying out the present LMTO-ASA based calculations, which provide a reasonable estimate of magnetic moments and energy differences\cite{oe}.  The IEC for Fe$_{3}$Nb$_{m}$($m$ = 1..16) configuration is then computed using Eq. (1), which is shown in Fig. 1. Rapid oscillations are observed up to 9 monolayers (ML) of Nb thickness, after which the oscillatory exchange coupling becomes appreciably weak. The coupling changes from ferromagnetic to antiferromagnetic configuration at about 2, 7 and 10 monolayers of Nb spacer, resulting in oscillation periods of 7.7 and 4.6 {\AA} respectively. Note that 1 ML corresponds to the interplanar thickness of 1.5335 {\AA}. 

To better understand the oscillation periods and the coupling phenomena in Fe/Nb multilayers, we fit the calculated variation of $J(m)$ with $m$ to the following asymptotic form\cite{mvs} 
\begin{widetext}
\begin{equation}
J(m) = \sum_{k=1}^{2}A_{k}sin\left(q_{k}m + \phi_{k}\right)/m^{2} + \sum_{l=3}^{4}A_{l}sin\left(q_{l}m + \phi_{l}\right)/m^{3},
\end{equation}
\end{widetext}
where $A_{k(l)}$'s describe the amplitudes; $q_{k(l)}$'s yield the periodicities $T_{k(l)}$(=2$\pi/q_{k(l)}$)'s; and $\phi_{k(l)}$'s the phases of the $k(l)$-th mode of oscillations. Our $ab$ $initio$ data for $J(m)$ in Eq. (1) yields a reasonably good fit to  Eq. (2) with the following four periods: $T_{1}$ =  4.14 ML ($\sim$  6.3 \AA), $T_{2}$ = 5.05 ML ($\sim$ 7.7 \AA), $T_{3}$ = 2.86 ML ($\sim$ 4.4 \AA),  $T_{4}$ =  20.28 ML ($\sim$ 31.1 \AA). These values fall well within the previous results\cite{matson,nitya} for the FeNb$_{m}$ heterostructures. Multiple periodicities arise due to existence of the multisheet Fermi surface in the Nb spacer layer. The Nb thickness periodicity of 7.7 {\AA} comes close to the experimental value of 9.0 {\AA}. This discrepancy is perhaps due to pre-asymptotic effects and the difficulty of including the true lattice structure in our calculations. Since in the present study, the total-energy calculations of the IEC are limited by 16 ML of spacer thickness, which might still be in the pre-asymptotic region, a slight discrepancy between the experimental result and total-energy calculation is expected. However, the 4.4 and 6.3 {\AA} periods are in fairly good agreement with the available $ab$ $initio$ data\cite{matson,nitya,sticht,stiles}.The well-fitted curve of Fig. 1 suggests that the interlayer exchange coupling in Fe/Nb multilayers has significant $1/m^{3}$ dependence in addition to the conventional $1/m^{2}$ dependence given by the RKKY theory. 

If we analyze the four periodicities that are obtained upon a reasonably good fitting of Eq. (2), the existence of higher harmonics is observed in the coupling function. These harmonics (denoted by $n$) add terms of the form $m^{-(2+i)}sin(2\pi nm/T + phase)$ with $i$ = 0, 1..., $n$ $\ge$ $2$, and $T$ being the fundamental period. This way the fourth harmonic ($n$ = 5) results in $m^{-2}sin(5\delta m + phase)$ with $\delta$ = $2\pi/T$. Now, if we express the 4.14 ML period $T_{1}$ as  $1/T_{1}$ = 1/5 + $\delta/2\pi$, the effective period\cite{mvs} is obtained as $T_{eff}$ = $2\pi/5\delta$ = $T_{1}/(5-T_{1})$ = 4.81 ML. This value is, however, a bit lower than the fit value of 5.05 ML, which corresponds to $T_{2}$. The slight discrepancy may be attributed to the uncertainties in the fit by Eq. (2). Expressing similarly the 2.86 ML period $T_{3}$ as $1/T_{3}$ = 1/3 + $\delta/2\pi$, the second harmonic ($n$ = 3) becomes $m^{-3}sin(3\delta m + phase)$, which yields an effective period $T^{'}_{eff}$ = $2\pi/3\delta$ = $T_{3}/(3-T_{3})$ = 20.43 ML. This value is also quite close to the fit value of $T_{4}$ (= 20.28 ML) with a discrepancy of about 0.7\%. Both the long periods $T_{2}$ and $T_{4}$ thus turn out to be the "Vernier" periods of the respective short period oscillations of $T_{1}$ and $T_{3}$. As pointed out by Schilfgaarde and Harrison\cite{mvs}, higher harmonics have significant presence in the quantum well limit where the RKKY description does not hold good. 

\subsection{Fermi surface and the RKKY periods}
In the RKKY approach, the oscillatory periods of the interlayer exchange coupling are uniquely determined by the stationary spanning vectors of the bulk Fermi surface of the spacer material. Several spanning vectors in the FS give rise to multiple periodicities, as we have already come across in the previous section. 

In Fe/Nb multilayers, the spacer layer Nb has five conduction electrons per atom that fill the first Brillouin zone completely while the second and third Brillouin zones partially\cite{karim}. The second zone is a closed "$octahedron$" (OCT), which contains a hole sheet centered at $\Gamma$. However, the third zone has two sheets. One sheet contains an open surface of holes, referred to as "$jungle$ $gym$" (JG), which extends from $\Gamma$ to $H$ points along the [100] direction. The other sheet is a set of distorted hole "$ellipsoids$" (ELL) centered at $N$ points. The OCT and JG sheets contact at three points\cite{matt} in the (100) and (110) symmetry planes of the Fermi surface of Nb. 

Simple square lattice planes of Fe and Nb with primitive translations of $\frac{\pi}{a}[100]$ and  $\frac{\pi}{a}[010]$ are stacked along the [001] growth direction to form Fe/Nb multilayers.  Since the translational symmetry in the growth direction is broken, multilayers behave like quasi-two-dimensional systems, which are periodic only in two dimensions\cite{man}. Hence, the Brillouin zone needs to be constructed in 2D to deal with the in-plane coordinates\cite{kol}. Fig. 2 shows the cross section of the spacer layer Fermi surface of Nb on the basis of our self-consistent calculations in the central (100) plane. The line $Q_{1}$, $Q_{2}$ and $Q_{3}$ as shown in Fig. 2, represent the spanning vectors for the [100] crystalline orientation; $Q_{1}$ spans the $\Gamma$-centered octahedron along the [100] direction, while $Q_{2}$ and  $Q_{3}$ span the outer and the inner ellipses respectively along the [100] direction. We find that the RKKY periods of 4.1, 6.4 and 7.5 {\AA}, as predicted from the Fermi surface spanning vectors, turn out to be in good agreement with the interlayer coupling periods of 4.4, 6.3 and 7.7 {\AA}, as  obtained by fitting the self-consistent results to Eq. (2). However, the Fermi surface topology of Nb is not able to predict the periodicity of 31.1 {\AA}. As we know from the preceding section, the long-wavelength coupling period of 31.1 {\AA} originates from the higher order terms in the coupling function of Eq. (2), which is also a higher harmonics of the short-wavelength period of 4.4 {\AA} . 

\subsection{Magnetization}
To examine how the Fe magnetic moment at the interface behaves as a function of the intervening layer thickness in Fe/Nb multilayers, we plot the magnetic moment of the interface monolayer of Fe vs. the spacer layer thickness of Nb in Fig. 3. There is a good agreement between our calculated results and the experimental observations as shown in the inset. We find that the Fe magnetic moment gets saturated in the asymptotic region with reduction of about 25 {\%} of the bulk value. This reduction in the measured data is about 40 {\%}. Following the works of Holmstr\"om $et$ $al$\cite{holm}, this kind of discrepancy may be attributed primarily to the interface alloying. 

The induced magnetic moments in the Nb spacer layer for a ferromagnetically ordered Fe$_{3}$Nb$_{16}$ superlattices are shown in Fig. 4(a). On the other hand, Fig. 4(b) displays the induced polarization for the antiferromagnetic configuration. Fe layers are at positions marked by 0 and 17 (not shown in the figure). The induced magnetic moment (in absolute values) in Nb spacer layer decreases from about 0.28 $\mu_{B}$ at the interface to about 0.02 $\mu_{B}$ further apart from the interface. For an Nb layer of 16 atomic planes, the calculated induced moment at the center of the Nb layer is about 10$^{-3}$ $\mu_{B}$. The period of oscillation of the induced moment in both the magnetic configurations turns out to be about 4.6 {\AA}, which is in fairly good agreement with the coupling period of 4.4 {\AA}, as determined by Eq. (2). The bias in the magnetic ordering of the induced moment (see Fig. 4), which occurs only in the ferromagnetic configuration, may be regarded as being due to the onset of non-RKKY terms in the coupling function when the Nb layer gets appreciably thicker. Mathon $et$ $al$\cite{mathon} have already shown analytically the presence of such non-RKKY terms for Co/Cu(001) system using the stationary phase approximation. 

However, to get a better understanding of the coupling behavior, one needs to examine the influence of Fe magnetization on the coupling strength. For this, we adopt a procedure similar to that of Schilfgaarde and Harrison\cite{mvs}. A trial density is constructed out of the charge densities of self-consistently calculated bulk ferromagnetic Fe and paramagnetic Nb in their respective atom-centered spheres. 
The charge density of Fe is constructed as follows\cite{mvs}
\begin{equation}
n_{Fe}(r) = n^{0}(r) \pm \alpha \frac{n^{\uparrow}(r) - n^{\downarrow}(r)}{2},
\end{equation}
where $n^{0}$ denotes the density of bulk paramagnetic Fe while $n^{\uparrow}$ and $n^{\downarrow}$ represent the spin densities of the majority and minority spins in ferromagnetic Fe and $\alpha$ is a parameter ranging from 0 to 1. The magnetic moment of Fe atom will be proportional to $\alpha$, taking full value at $\alpha$ = 1 and 0 value when $\alpha$ = 0.
Since we are interested in obtaining $J$ as a function of the Fe moment, we have used the trial densities in our frozen-potential calculations pertaining to relevant supercells. To have a preliminary idea of how the coupling and the moments are inter-related in Fe/Nb systems, we construct a 16-atom Nb supercell with two Fe atoms substituted such that one Fe sits at (0 0 0) while the other at 3a(1 1 1)/2. For parallel and antiparallel alignment of two inequivalent Fe atoms, the energy difference is calculated\cite{fn1} as a function of the Fe moment parametrized by $\alpha$. According to the RKKY theory, this energy difference should vary as $\alpha^{2}$ (see Fig. 5).  

Several superlattices out of Fe$_{3}$Nb$_{m}$ (m = 1--16) multilayer configuration are constructed subsequently to calculate the energy difference $J_{\alpha}$($m$) = $E$[Fe$_{3}^{\uparrow}$Nb$_{m}$Fe$_{3}^{\downarrow}$Nb$_{m}$] - $E$[Fe$_{3}^{\uparrow}$Nb$_{m}$Fe$_{3}^{\uparrow}$Nb$_{m}$], as function of $\alpha$. We determine the amplitudes A$_{k(l)}$'s for each $\alpha$ by fitting $J_{\alpha}$($m$) to the functional form as given by Eq. (2). This way, we find two short-period amplitudes and two long-period amplitudes, which are illustrated in Fig 5. In Fe/Cr multilayers, Schilfgaarde and Harrison\cite{mvs} found  the coupling amplitudes (for small $\alpha$) of both short and long periods to increase as $\alpha^{2}$, while at larger $\alpha$, short period amplitude continued its rise, though the long period saturated completely. Here we find that the initial $\alpha^{2}$ dependence of the exchange coupling, as assumed in the RKKY theory, at small $\alpha$ values is followed by two distinct features at higher $\alpha$ values viz. \\
(i) stronger dependence on the Fe moment as IEC shoots up for short period oscillations and \\
(ii) complete independence of the Fe moment as IEC gets saturated for long period oscillations. \\ 

As a function of magnetization ($\mf$), the interlayer exchange energy can be described as\cite{fuller}
\begin{equation}
J_{\mf} = J_{1}(\mbf.\mcf) + J_{2}(\mbf.\mcf)^{2},
\end{equation}
where $J_{1}$ and $J_{2}$ are respectively the bilinear and biquadratic coupling constants, while $\mbf$ and $\mcf$ denote the magnetizations of the adjacent ferromagnetic layers. In order to fit the calculated results for the 16-atom supercell data, we need to expand the above expression in the following form     
\begin{equation}
J_{\alpha} = J_{1}{\alpha}^{2} + J_{2}{\alpha}^{4} + J_{3}{\alpha}^{6},
\end{equation}
where $\mf$ has been parametrized by $\alpha$ and $J_{3}$ is an additional triquadratic term appearing in the exchange coupling. Initially, we calculate the coupling energy, as a function of  $\alpha$, for the 16-atom Nb supercell with only two Fe atoms substituted. These are represented by asterisks in Fig. 5. While fitting the data we find that the coupling energy has biquadratic and triquadratic terms in addition to bilinear terms in $\alpha$. In doing so, the coupling constants turn out to be in the ratio of about 1:2:4. It may be noted that the RKKY theory assumes only the bilinear terms in $\alpha$. The linear part of the fitted curve in Fig. 5 thus shows the RKKY kind of coupling in the region of small magnetic moment of Fe.
We then calculate the interlayer exchange coupling for the Fe$_{3}$Nb$_{m}$ (m = 1--16) multilayer configuration as a function of $\alpha$.
Short period amplitudes of the Fe$_{3}$Nb$_{m}$ (m = 1--16) multilayer configuration are found to follow the above phenomenological expression for the interlayer interaction, but the long period amplitudes show a saturation behavior at large $\alpha$. For a complete confinement, the coupling becomes independent of the size of the magnetic moment\cite{dme}, as demonstrated by the long period amplitudes. The saturation of the long-period amplitudes with the ferromagnetic Fe moment thus favors  a QW description of the exchange coupling. Because of the incomplete confinement of states in Nb, Fe/Nb multilayers exhibit a partial transition to the QW character from the RKKY description, which is valid at small moments [Fig. 5]. 

\subsection{Energy bands, density of states and the quantum well model}
To study the nature of the electronic states in Fe/Nb multilayers, we resort to the bulk band structures (see Fig. 6) of the ferromagnetic Fe (for both spins) and the paramagnetic Nb along the [100] as well as the [110] direction. These directions are important\cite{gwc}  for  multiple scattering, since the respective $\Delta_{2}$, $\Sigma_{1}$ and $\Sigma_{2}$ bands cross the Fermi level (E$_{F}$) in the spacer medium contributing to the oscillations in the interlayer coupling. 

The Fe minority $\Gamma_{12}$-$\Delta_{2}$-H$_{12}$ band closely resembles the corresponding Nb band indicating that the Fe/Nb interface will be more transparent to the minority-spin electrons  than to the majority-spin electrons. 
The vanishing overlap between the majority spin bands $\Gamma_{12}$-$\Delta_{2}$-$H_{12}$ in Nb and Fe creates spin-dependent gaps at the Fe/Nb interface resulting in energy barriers that confine electrons to the intervening Nb layer. The electrons in Nb  band $\Gamma_{12}$-$\Delta_{2}(\uparrow)$-$H_{12}$ can not pass into the Fe majority  band since there are no available states of $\Delta_{2}(\uparrow)$ symmetry from $\Gamma_{12}$ (Fe$\uparrow$) to $\Gamma_{12}$ (Nb). As a result, spin-up electrons with energies from (E$_{F}$ - 0.14) eV to (E$_{F}$ + 3.15) eV experience multiple reflections at the interface and get confined to the Nb spacer. Similar phenomena happen along the [110] direction. The electrons in the Nb band $\Gamma_{25'}$-$\Sigma_{1}(\uparrow)$-$N_{1'}$ can not pass into the Fe majority band in the energy range from (E$_{F}$ - 0.94) eV to (E$_{F}$ + 2.38) eV. Similarly, the Nb electrons in $\Gamma_{25'}$-$\Sigma_{2}(\uparrow)$-$N_{2}$ band can not pass into the Fe majority band in the energy range from (E$_{F}$ - 2.29) eV to (E$_{F}$ + 0.56) eV. The confinement of electrons thus occurs in the spacer layer resulting in the formation of quantum well states. As Fig. 6 shows, $bcc$ Fe has mostly minority-spin states at the Fermi level and this gives rise to majority-spin quantum well states in Nb. 

Fig. 7 displays how the density of states at $E_{F}$ oscillates with the spacer layer thickness, which is a characteristic feature of the quantum well model. According to the QW model, the oscillatory behavior of the magnetic coupling in Fe/Nb multilayers as demonstrated by Fig. 1  can be traced back to the oscillations of the density of states (DOS) at the Fermi level\cite{dm,stiles}. During the formation of QW states, Fe/Nb interfaces act as electron mirrors that induce standing waves in the Nb spacer medium. As Fig. 7 suggests, the first two maxima correspond to the  oscillation period of about 4.6 {\AA}. The periodicity then enhances to about 6.1 {\AA}. We notice that the QW periodicities of 4.6 and 6.1 {\AA} are in fairly good agreement with the  interlayer coupling periods of 4.4 and 6.3 {\AA} as obtained from the self-consistent results for Fe$_{3}$Nb$_{m}$ (m=1..16) multilayers (see Sec. III-A). It implies that there is a reasonable connection between the existence of the spin-polarized QW states and the manifestation of magnetic coupling in Fe/Nb multilayers. 

Oscillations in the density of states at the Fermi level can also be explained on the basis of the bulk band structures of Fe and Nb, assuming the QW model. We know from the previous discussion that the majority-spin energy bands of Fe along the [100] and the [110] directions provide the confinement to Nb electrons in forming the QW states inside the spacer medium. Fig. 6 shows that $\Delta_{2}$ and $\Sigma_{1}$ bands lead to the formation of QW states around $E_{F}$ inside the Nb layer. It is because there are no propagating states of the same symmetry at the same energy associated with the majority-spin electrons upon crossing the Fermi level. We obtain a band offset of 4.1 eV for the $\Delta_{2}$ band in the [100] direction, and another band offset of 3.3 eV for the $\Sigma_{1}$ band in the [110] direction [see Fig. 6]. Ortega $et$ $al$\cite{himp3} have shown for Co/Cu and Fe/Cu systems that a large band offset between the band edges of the magnetic and the spacer layer gives rise to sharply confined wavefunctions.

For $z$ being the multilayer growth direction, the translational symmetry holds only in the $x$-$y$ plane so that 
the wavefunction of an electron in a quantum well can be expressed as\cite{mun}
\begin{equation}
\psi_{\kbf,k}(\Rbf) = Af_{k}(z)e^{i{\kbf.\rbf}}u_{\kbf,k}(\Rbf),
\end{equation}
where $A$ is a normalization constant; $\kbf$ = ($kx$,$ky$); $\rbf$ = ($x$,$y$); $u_{\kbf,k}(\Rbf)$ is periodic in the lattice while $f_{k}(z)$ is the envelope function, which ensures that the boundary conditions are met at the interfaces. Eq. (6) shows that the wavefunction of a QW state consists of a rapidly oscillating Bloch function, which is modulated by a slowly varying envelope function. For a single band edge, the modulation of the Bloch wave ($k_{edge}$) by an envelope ($k_{env}$) yields a total wavevector\cite{himp3}
\begin{equation}
k_{tot} = k_{edge} \pm k_{env},
\end{equation} 
where only the normal components of the wavevectors are considered since the boundary conditions for the components that are parallel to the interface mimic the bulk. By drawing an analogy with the simple interferometer model, we may expect the interference maxima to appear at every half wavelength ($\lambda_{env}$) of the envelope as the thickness of the Nb spacer layer is increased so that QW states appear with a period of $\lambda_{env}/2$. For QW states at the Fermi level, the total wavevector becomes the Fermi wavevector if the spacer layer is thick enough to exhibit bulk-like bonding\cite{himp1}. As Fig. 6(a) shows, the band edge for the $\Sigma_{1}$ band in the [110] direction is located at the zone boundary ($k_{BZ}$) so that $k_{env}$ = $(k_{BZ}$ - $k_{F})$. The frequency associated with this $k_{env}$ in real space may be viewed as a beat frequency between the Fermi wavelength and the atom spacing\cite{himp2}. The oscillation period thus becomes 
\begin{equation}
 T = \pi/k_{env} = \pi/(k_{BZ} - k_{F})
\end{equation}
This way, we obtain a period of 5.8 {\AA} due to the $\Sigma_{1}$ band, which is in fairly good agreement with the   periodicity of 6.1 {\AA}, as appeared in the oscillating DOS at $E_{F}$ (see Fig. 7). It may be noted that the wavevector determining the periodicity of the envelope function along the [110] direction is quite small and thus the wavelength is large. However, a reverse situation occurs for the $\Delta_{2}$ band in the [100] direction, where 
$(k_{BZ}$ - $k_{F})$ is large as $k_{F}$ is small. The envelope function should produce in this case a short-wavelength weak modulation of the QW wavevector\cite{mun,smith}. 

\subsection{Phase accumulation model}
A simple way of predicting the energy for QW states is the $phase$ $accumulation$ $model$ (PAM), which is often used to describe the thickness dependence of QW energies and resonances in thin films and other layered structures\cite{smith}. According to this model, the condition for a QW state is determined by summing over all the phases accumulated by a propagating plane wave inside a quantum well. The total phase accumulated must be an integral multiple of 2$\pi$. 
For $m$ monolayers of Nb spacer, each of thickness $d$, the total spacer width becomes $d_{Nb}$ = $md$. Thus the distance traversed by an electron in the spacer medium is 2$md$ resulting in 2$mdk^{\perp}$ phase change, where $k^{\perp}$ represents the electron wavevector normal to the layers. If the phase shift of the electron wavefunction upon reflection at each interface is $\phi_{I}$, the total phase in a round trip within the spacer medium can be written as\cite{smith}
\begin{equation}
2k^{\perp}d_{Nb} + 2\phi_{I}(E) = 2n\pi,
\end{equation}
where $n$ is an integer related to the number of half-wavelengths that span the quantum well. 

The phase shift at each interface in Eq. (9) can be calculated by making use of following $ansatz$\cite{smith}
\begin{equation}
\phi_{I}(E) = 2\sin^{-1}\sqrt{\frac{E-E_{L}}{E_{U}-E_{L}}}-\pi,
\end{equation}  
where $E_{U}$ and $E_{L}$ represent the energies of the potential well taken from the upper and lower edges of the energy gap in the Fe majority-spin band, which acts as a potential energy barrier for the propagating electrons. The interface reflection phase evolves across a band gap from 0 to -$\pi$ in traversing from the top to the bottom of the potential well.

In order to obtain $k^{\perp}$ in Eq. (9), we make use of a two-band nearly free electron (NFE) model, which approximates the Nb electrons in the $\Delta_{2}$ and $\Sigma_{1}$ bands. According to this model, the wavevector of a NFE band can be expressed as\cite{qiu,alt}
\begin{widetext}
\begin{equation}
k^{\perp}/k_{BZ} = 1 - \left[1 + (E - V_{0})/G - \sqrt{4(E - V_{0})/G + (U/G)^{2}} \right]^{1/2},
\end{equation}
\end{widetext}
where $G$ = $\hbar^{2}k_{BZ}^{2}/2m^{*}$, where $m^{*}$ is the electron effective mass; $V_{0}$ is a constant offset of the periodic potential; 2$U$ is the energy gap at the zone boundary; and $E$ is the electron energy with respect to the Fermi level. Eq. (10) thus needs three parameters to determine $k^{\perp}$ for a given energy $E$. A fit to the self-consistently calculated $\Delta_{2}$ band of Nb yields $U$ = 2.05 eV, $V_{0}$ = -9.85 eV and $m^{*}$ = 1.08$m_{e}$, where $m_{e}$ is the electron mass. On the other hand, upon fitting Eq. (11) to the $\Sigma_{1}$ band of Nb, we have $m^{*}$ = 1.05$m_{e}$, $U$ = 1.66 eV and $V_{0}$ = -5.5 eV.

The presence of a periodic atomic potential in the spacer layer, however, ensures that in addition to two traveling waves with wavevectors $k$ and -$k$ arising from reflections at the interfaces, electron wavefunctions in the phase accumulation model should be described\cite{smith, mil} by the combination of two more waves with wavevectors ($k$-$g$) and -($k$-$g$), where $g$ is the reciprocal lattice vector. These additional waves correspond to the Bragg scattering within the periodic potential of the spacer layer. Since the reciprocal lattice vector normal to the interface within the first Brillouin zone is $g$ = 2$k_{BZ}$ = 2$\pi/d$, we have 2$k_{BZ}d_{Nb}$ = 2$m\pi$. Using this relation, Eq. (9) can be rewritten as  
\begin{equation}
2\kappa^{\perp}d_{Nb} - 2\phi_{I}(E) = 2\nu\pi,
\end{equation}
where $\kappa^{\perp}$ = $k_{BZ}$ - $k^{\perp}$ and $\nu$ = $m$ - $n$. It follows from Eq. (8) of the preceding section that $\kappa^{\perp}$ is characteristic of the envelope function that modulates the QW wavefunction. 

As we know, states within Nb layers that do not coincide in energy and momentum with those within Fe layers form the quantum well states. With the increase in the Nb layer thickness, the positions of the spin polarized QW states vary and thus exhibit regular dispersion through the Fermi level.  For $n$ = 1, Figs. 8(a) and 8(b) demonstrate the nature of QW dispersions in Fe/Nb multilayers along [100] and [110] directions as generated by Eq. (12) of the phase accumulation model. We find that the $\Delta_{2}$ band yields weak dispersion as compared to the $\Sigma_{1}$ band of Nb due mainly to the weak modulation of $k^{\perp}$ in the [100] direction. The oscillation periods of the quantum well states at the Fermi energy, as shown by the dispersion curves of 8(a) and 8(b), are obtained as about 5.5 and 7.3 {\AA}. These values are close to the fitted values of 6.3 and 7.7  {\AA} respectively. 
 
\section{Conclusions}
In this paper, we have explained the phenomena of interlayer exchange coupling in Fe/Nb(001) multilayers in terms of RKKY as well as the QW model using the density functional calculations. The RKKY periodicities arising out of extremal spanning vectors of the multisheet Nb Fermi surface have been found to be in favorable agreement with the available experimental as well as calculated data. In the region of small magnetic moments of Fe, both the short and long periods display the RKKY kind of coupling, since IEC shows bilinearity in magnetization. However, the presence of additional biquadratic and triquadratic coupling constants in IEC at higher Fe moments signify the onset of the $non$-RKKY character, especially of the short-period oscillations. We have found that at moments closer to the bulk value, the long-period oscillations become eventually independent of the size of the magnetic moments. The appearance of higher harmonics in the well-fitted envelope of the oscillatory exchange coupling turns out to be responsible for the saturation of long periods. The oscillatory nature of the density of states at the Fermi level supports the QW description of the exchange coupling in Fe/Nb multilayers. It is because such oscillations originate mainly from the spin-dependent confinement of the propagating electrons inside the quantum well of the spacer medium. Quantum well dispersions around the Fermi level illustrate that the majority spin bands contribute largely to the formation of QW states, which is subsequently analyzed by making use of the $phase$ $accumulation$ $model$. All these results show that magnetic quantum wells are formed in Fe/Nb multilayers and the oscillatory behavior of the exchange coupling in Fe/Nb multilayers is better described by the Quantum Well model.

\begin{acknowledgments}
It's a pleasure to thank Profs. P. Bruno, T. V. Ramakrishnan, D. Kumar, 
M. K. Harbola and R. C. Budhani for helpful discussions. This work was supported by the Asian Office of Aerospace Research and Development via contract No. AOARD-06-4019.
\end{acknowledgments}

\begin{figure}[h]
{\bf Figure Captions}
\caption{\label{fig1}The calculated oscillatory interlayer exchange coupling (solid circles) as a function of the number of Nb spacer layers in Fe$_{3}$Nb$_{m}$($m$ = 1--16) multilayers. The solid line is the fitted plot (see the text for details).The thickness of 1 Nb layer=1.5335 {\AA}.}

\caption{\label{fig2}(Color online) Cross sections of the Fermi surface of Nb in the (100) plane. $\Gamma$ labels the center of the Brillouin zone, $N$ indicates the center of each face of the dodecahedron and $H$ labels the corners of the four-fold symmetry on the zone boundary.}

\caption{\label{fig3}Variation of the Fe magnetic moment with Nb spacer thickness in Fe$_{3}$Nb$_{m}$($m$ = 1--16) heterostructures (see the text for details). The inset shows the experimental results by Mattson et al\cite{matson}.}

\caption{\label{fig4}The induced magnetic moment in Nb spacer layers for Fe$_{3}$Nb$_{16}$ heterostructures in the (a) ferromagnetic (FM) and (b) antiferromagnetic (AFM) orientations. The thickness of 1 Nb layer=1.5335 {\AA}.}

\caption{\label{fig5}(Color online) Interlayer exchange coupling in Fe/Nb multilayers as a function of Fe magnetic moment parametrized by $\alpha$ (see the text for details).The asterisks represent the results for the 16-atom supercell while the solid line depicts the fitted curve according to Eq. (5). Note that the linear part of the solid line resembles the RKKY kind of coupling. The triangles and diamonds indicate the short-period amplitudes while the circles and squares represent the long-period amplitudes of the Fe$_{3}$Nb$_{m}$($m$ = 1--16) multilayer system.}

\caption{\label{fig6}The bulk energy bands of Nb and $bcc$ Fe ($\uparrow$ and $\downarrow$) along  the (a) [100] and (b) [110] directions. The bands with $\Delta_{2}$ and $\Sigma_{1}$ symmetries are displayed by solid lines while the bands with $\Sigma_{1}$ symmetry by dashed lines. 
 Only majority-spin states in Nb exhibit quantum well character at the Fermi level since the minority-spin  $\Delta_{2}$ and $\Sigma_{1}$ states couple with the corresponding states in Fe.}

\caption{\label{fig7}Oscillations in the density of states at the Fermi level, $E_{F}$, with the Nb spacer thickness, caused by the quantum well states in Fe$_{3}$Nb$_{m}$($m$ = 1--16) heterostructures.} 

\caption{\label{fig8}The thickness dependence of the QW energies in Fe/Nb multilayers generated by Eq. (12) of the phase accumulation model with respect to (a) $\Delta_{2}$ and (b) $\Sigma_{1}$ bands.}

\end{figure}


\begin{thebibliography}{<num>}
\bibitem{fn1}We tested our calculations by carrying out the exercise for the 54-atom Fe/Cr supercell. We obtained $\alpha^{2}$ behavior in the $J$ vs. $\alpha$ plot, which is in agreement with Ref. 8.
\end{thebibliography}
\end{document}